\documentstyle[floats,prd,aps,epsf]{revtex}
\draft 

\begin{document}


\twocolumn[\hsize\textwidth\columnwidth\hsize\csname
@twocolumnfalse\endcsname

\title{On the Numerical Integration of Einstein's Field Equations}

\author{Thomas W.~Baumgarte$^1$ and Stuart L.~Shapiro$^2$}

\address{${}^1$ Department of Physics, University of Illinois at
	Urbana-Champaign, Urbana, Il~61801}

\address{${}^2$ Departments of Physics and Astronomy and NCSA, 
	University of Illinois at Urbana-Champaign, Urbana, Il~61801}

\maketitle

\begin{abstract}
Many numerical codes now under development to solve Einstein's
equations of general relativity in 3+1 dimensional spacetimes employ
the standard ADM form of the field equations.  This form involves
evolution equations for the raw spatial metric and extrinsic curvature
tensors.  Following Shibata and Nakamura, we modify these equations by
factoring out the conformal factor and introducing three ``connection
functions''.  The evolution equations can then be reduced to wave
equations for the conformal metric components, which are coupled to
evolution equations for the connection functions.  We evolve small
amplitude gravitational waves and make a direct comparison of the
numerical performance of the modified equations with the standard ADM
equations.  We find that the modified form exhibits much improved
stability.
\end{abstract}


\vskip2pc]


\section{INTRODUCTION}

The physics of compact objects is entering a particularly exciting
phase, as new instruments can now yield unprecedented observations.
For example, there is evidence that the Rossi X-ray Timing Explorer
has identified the innermost stable circular orbit around an accreting
neutron star~\cite{zsss98}.  Also, the new generation of gravitational
wave detectors under construction, including LIGO, VIRGO, GEO and
TAMA, promise to detect, for the first time, gravitational radiation
directly (see, e.g.,~\cite{t95}).

In order to learn from these observations (and, in the case of the
gravitational wave detectors, to dramatically increase the likelihood
of detection), one has to predict the observed signal from theoretical
modeling.  The most promising candidates for detection by the
gravitational wave laser interferometers are the coalescences of black
hole and neutron star binaries.  Simulating such mergers requires
self-consistent, numerical solutions to Einstein's field equations in
3 spatial dimensions, which is extremely challenging.  While several
groups, including two ``Grand Challenge Alliances''~\cite{gc}, have
launched efforts to simulate the coalescence of compact objects (see
also~\cite{on97,wmm96}), the problem is far from being solved.

Before Einstein's field equations can be solved numerically, they have
to be cast into a suitable initial value form.  Most commonly, this is
done via the standard 3+1 decomposition of Arnowitt, Deser and Misner
(ADM,~\cite{adm62}). In this formulation, the gravitational fields are
described in terms of spatial quantities (the spatial metric and the
extrinsic curvature), which satisfy some initial constraints and can
then be integrated forward in time.  The resulting ``$\dot g - \dot
K$'' equations are straightforward, but do not satisfy any known
hyperbolicity condition, which, as it has been argued, may cause
stability problems in numerical implementations.  Therefore, several
alternative, hyperbolic formulations of Einstein's equations have been
proposed~\cite{fr94,bmss95,aacy96,pe96,f96,acy98}.  Most of these
formulations, however, also have disadvantages.  Several of them
introduce a large number of new, first order variables, which take up
large amounts of memory in numerical applications and require many
additional equations. Some of these formulations require taking
derivatives of the original equations, which may introduce further
inaccuracies, in particular if matter sources are present.  It has
been widely debated if such hyperbolic formulations have computational
advantages~\cite{texas95}; their performance has yet to be compared
directly with that of the original ADM equations.  Accordingly, it is
not yet clear if or how much the numerical behavior of the ADM
equations suffers from their non-hyperbolicity.

In this paper, we demonstrate by means of a numerical experiment and a
direct comparison that the standard implementation of the ADM system
of equations, consisting of evolution equations for the bare metric
and extrinsic curvature variables, is more susceptible to numerical
instabilities than a modified form of the equations based on a
conformal decomposition as suggested by Shibata and
Nakamura~\cite{sn95}.  We will refer to the standard, ``$\dot g - \dot
K$'' form of the equations as ``System I'' (see Section~\ref{sys1}
below).  We follow Shibata and Nakamura and modify these original ADM
equations by factoring out a conformal factor and introducing a
spatial field of connection functions (``System II'', see
Section~\ref{sys2} below).  The conformal decomposition separates
``radiative'' variables from ``nonradiative'' ones in the spirit of
the ``York-Lichnerowicz'' split~\cite{l44,y71}.  With the help of the
connection functions, the Ricci tensor becomes an elliptic operator
acting on the components of the conformal metric.  The evolution
equations can therefore be reduced to a set of wave equations for the
conformal metric components, which are coupled to the evolution
equations for the connection functions.  These wave equations
reflect the hyperbolic nature of general relativity, and can also be
implemented numerically in a straight-forward and stable manner.

We evolve low amplitude gravitational waves in pure vacuum spacetimes,
and directly compare Systems I and II for both geodesic slicing and
harmonic slicing.  We find that System II is not only more appealing
mathematically, but performs far better numerically than System I.  In
particular, we can evolve low amplitude waves in a stable fashion for
hundreds of light travel timescales with System II, while the
evolution crashes at an early time in System I, independent of gauge
choice.  We present these results in part to alert developers of 3+1
general relativity codes, many of whom currently employ System I, that
a better set of equations may exist for numerical implementation.

The paper is organized as follows.  In Section~\ref{sec2}, we present
the basic equations of both Systems I and II.  We briefly discuss
our numerical implementation in Section~\ref{sec3}, and present
numerical results in Section~\ref{sec4}.  In Section~\ref{sec5}, we
summarize and discuss some of the implications of our findings.


\section{BASIC EQUATIONS}
\label{sec2}

\subsection{System I}
\label{sys1}

We write the metric in the form
\begin{equation}
ds^2 = - \alpha^2 dt^2 + \gamma_{ij} (dx^i + \beta^i dt)(dx^j + \beta^j dt),
\end{equation}
where $\alpha$ is the lapse function, $\beta^i$ is the shift vector, and
$\gamma_{ij}$ is the spatial metric.  Throughout this paper, Latin indices
are spatial indices and run from 1 to 3, whereas Greek indices are spacetime
indices and run from 0 to 3.  The extrinsic curvature $K_{ij}$
can be defined by the equation
\begin{equation} \label{gdot1}
\frac{d}{dt} \gamma_{ij} = - 2 \alpha K_{ij},
\end{equation}
where 
\begin{equation}
\frac{d}{dt} = \frac{\partial}{\partial t} - {\cal L}_{\beta}
\end{equation}
and where ${\cal L}_{\beta}$ denotes the Lie derivative with respect to
$\beta^i$.

The Einstein equations can then be split into the Hamiltonian constraint
\begin{equation} \label{ham1}
R - K_{ij}K^{ij} + K^2 = 2 \rho,
\end{equation}
the momentum constraint
\begin{equation} \label{mom1}
D_j K^{j}_{~i} - D_i K = S_i,
\end{equation}
and the evolution equation for the extrinsic curvature
\begin{equation} \label{Kdot1}
\frac{d}{dt} K_{ij} = - D_i D_j \alpha + \alpha ( R_{ij} 
        - 2 K_{il} K^l_{~j} + K K_{ij} - M_{ij} ) 
\end{equation}
Here $D_i$ is the covariant derivative associated with $\gamma_{ij}$,
$R_{ij}$ is the three-dimensional Ricci tensor
\begin{eqnarray} \label{ricci}
R_{ij} & = & \frac{1}{2} \gamma^{kl} 
	\Big( \gamma_{kj,il} + \gamma_{il,kj} 
		- \gamma_{kl,ij} - \gamma_{ij,kl} \Big) \\[1mm]
 	& & + \gamma^{kl} \Big( \Gamma^m_{il} \Gamma_{mkj}
 	- \Gamma^m_{ij} \Gamma_{mkl} \Big), \nonumber
\end{eqnarray}
and $R$ is its trace $R = \gamma^{ij} R_{ij}$.  We have also introduced the
matter sources $\rho$, $S_i$ and $S_{ij}$, which are projections of
the stress-energy tensor with respect to the unit normal vector
$n_{\alpha}$,
\begin{eqnarray}
\rho & = & n_{\alpha} n_{\beta} T^{\alpha \beta}, \nonumber \\[1mm]
S_i  & = & - \gamma_{i\alpha} n_{\beta} T^{\alpha \beta}, \\[1mm]
S_{ij} & = & \gamma_{i \alpha} \gamma_{j \beta} T^{\alpha \beta}, \nonumber
\end{eqnarray}
and have abbreviated
\begin{equation}
M_{ij} \equiv S_{ij} + \frac{1}{2} \gamma_{ij}(\rho - S),
\end{equation}
where $S$ is the trace of $S_{ij}$, $S = \gamma^{ij} S_{ij}$.

The evolution equations~(\ref{gdot1}) and~(\ref{Kdot1}) together with
the constraint equations~(\ref{ham1}) and~(\ref{mom1}) are equivalent
to the Einstein equations, and are commonly referred to as the ADM
form of the gravitational field equations~\cite{adm62,footnote1}.  We
will call these equations System I.  This system is widely used in
numerical relativity calculations (e.g.~\cite{aetal98,cetal98}), even
though its mathematical structure is not simple to characterize and
may not be ideal for computation.  In particular, the Ricci
tensor~(\ref{ricci}) is not an elliptic operator: while the last one
of the four terms involving second derivatives,
$\gamma^{kl}\gamma_{ij,kl}$, is an elliptic operator acting on the
components of the metric, the elliptic nature of the whole operator is
spoiled by the other three terms involving second derivatives.
Accordingly, the system as a whole does not satify any known
hyperbolicity condition (see also the discussion in~\cite{f96}).
Therefore, to establish existence and uniqueness of solutions to
Einstein's equations, most mathematical analyses rely either on
particular coordinate choices or on different formulations.

\subsection{System II}
\label{sys2}

Instead of evolving the metric $\gamma_{ij}$ and the extrinsic
curvature $K_{ij}$, we can evolve a conformal factor and the trace
of the extrinsic curvature separately (``York-Lichnerowicz
split''~\cite{l44,y71}).  Such a split is very appealing from both a
theoretical and computational point of view, and has been widely
applied in numerical axisymmetric (2+1) calculations (see,
e.g.,~\cite{e84}).  More recently, Shibata and Nakamura~\cite{sn95}
applied a similar technique in a three-dimensional (3+1) calculation. 
Adopting their notation, we write the conformal metric as
\begin{equation}
\tilde \gamma_{ij} = e^{- 4 \phi} \gamma_{ij}
\end{equation}
and choose
\begin{equation}
e^{4 \phi} = \gamma^{1/3} \equiv \det(\gamma_{ij})^{1/3},
\end{equation}
so that the determinant of $\tilde \gamma_{ij}$ is unity. We also 
write the trace-free part of the extrinsic curvature $K_{ij}$ as
\begin{equation}
A_{ij} = K_{ij} - \frac{1}{3} \gamma_{ij} K,
\end{equation}
where $K = \gamma^{ij} K_{ij}$. It turns out to be convenient to introduce
\begin{equation}
\tilde A_{ij} = e^{- 4 \phi} A_{ij}.
\end{equation}
We will raise and lower indices of $\tilde A_{ij}$ with the conformal
metric $\tilde \gamma_{ij}$, so that
$\tilde A^{ij} = e^{4 \phi} A^{ij}$ (see~\cite{sn95}).

Taking the trace of the evolution equations~(\ref{gdot1})
and~(\ref{Kdot1}) with respect to the physical metric $\gamma_{ij}$,
we find~\cite{footnote2}
\begin{equation} \label{phidot2}
\frac{d}{dt} \phi = - \frac{1}{6} \alpha K
\end{equation}
and
\begin{equation} \label{Kdot2}
\frac{d}{dt} K = - \gamma^{ij} D_j D_i \alpha + 
	\alpha(\tilde A_{ij} \tilde A^{ij}
	+ \frac{1}{3} K^2) + \frac{1}{2} \alpha (\rho + S),
\end{equation}
where we have used the Hamiltonian constraint~(\ref{ham1}) to
eliminate the Ricci scalar from the last equation.  The tracefree
parts of the two evolution equations yield
\begin{equation} \label{gdot2}
\frac{d}{dt} \tilde \gamma_{ij} =  
	- 2 \alpha \tilde A_{ij}.
\end{equation}
and
\begin{eqnarray} \label{Adot2}
\frac{d}{dt} \tilde A_{ij} & = & e^{- 4 \phi} \left( 
	- ( D_i D_j \alpha )^{TF}  +
	\alpha ( R_{ij}^{TF} - S_{ij}^{TF} ) \right)
	\nonumber \\[1mm]
	& & + \alpha (K \tilde A_{ij} - 2 \tilde A_{il} \tilde A^l_{~j}).
\end{eqnarray}
In the last equation, the superscript $TF$ denotes the trace-free part
of a tensor, e.g. $R_{ij}^{TF} = R_{ij} - \gamma_{ij} R/3$.  Note that
the trace $R$ could again be eliminated with the Hamiltonian
constraint~(\ref{ham1}).  Note also that $\tilde \gamma_{ij}$ and
$\tilde A_{ij}$ are tensor densities of weight $-2/3$, so that their
Lie derivative is, for example,
\begin{equation}
{\cal L}_{\beta} \tilde A_{ij} = \beta^k \partial_k \tilde A_{ij}
	+ \tilde A_{ik} \partial_j \beta^k 
	+ \tilde A_{kj} \partial_i \beta^k
	- \frac{2}{3} \tilde A_{ij} \partial_k \beta^k.
\end{equation}

The Ricci tensor $R_{ij}$ in~(\ref{Adot2}) can be written as the sum
\begin{equation}
R_{ij} = \tilde R_{ij} + R_{ij}^{\phi}.
\end{equation}
Here $R_{ij}^{\phi}$ is
\begin{eqnarray}
R^{\phi}_{ij} & = & - 2 \tilde D_i \tilde D_j \phi - 
	2 \tilde \gamma_{ij} \tilde D^l \tilde D_l \phi \nonumber \\[1mm]
	& & + 4 (\tilde D_i \phi)(\tilde D_j \phi)
	- 4 \tilde \gamma_{ij} (\tilde D^l \phi) (\tilde D_l \phi),
\end{eqnarray}
where $\tilde D_i$ is the derivative operator associated with $\tilde
\gamma_{ij}$, and $\tilde D^i = \tilde \gamma^{ij} \tilde D_j$.

The ``tilde'' Ricci tensor $\tilde R_{ij}$ is the Ricci tensor
associated with $\tilde \gamma_{ij}$, and could be computed by
inserting $\tilde \gamma_{ij}$ into equation~(\ref{ricci}).
However, we can bring the Ricci tensor into a manifestly elliptic
form by introducing the ``conformal connection functions''
\begin{equation} \label{cgsf}
\tilde \Gamma^i \equiv \tilde \gamma^{jk} \tilde \Gamma^{i}_{jk}
	= - \tilde \gamma^{ij}_{~~,j},
\end{equation}
where the $\tilde \Gamma^{i}_{jk}$ are the connection coefficients
associated with $\tilde \gamma_{ij}$, and where the last equality
holds because $\tilde \gamma = 1$.  In terms of
these, the Ricci tensor can be written~\cite{footnote3}
\begin{eqnarray}
\tilde R_{ij} & = & - \frac{1}{2} \tilde \gamma^{lm}
	\tilde \gamma_{ij,lm} 
	+ \tilde \gamma_{k(i} \partial_{j)} \tilde \Gamma^k
	+ \tilde \Gamma^k \tilde \Gamma_{(ij)k}  + \nonumber \\[1mm]
	& & \tilde \gamma^{lm} \left( 2 \tilde \Gamma^k_{l(i} 
	\tilde \Gamma_{j)km} + \tilde \Gamma^k_{im} \tilde \Gamma_{klj} 
	\right).
\end{eqnarray}	
The principal part of this operator, $\tilde \gamma^{lm} \tilde
\gamma_{ij,lm}$, is that of a Laplace operator acting on the
components of the metric $\tilde \gamma_{ij}$.  It is obviously
elliptic and diagonally dominant (as long as the metric is diagonally
dominant).  All the other second derivatives of the metric appearing
in~(\ref{ricci}) have been absorbed in the derivatives of the
connection functions.  At least in appropriately chosen coordinate
systems (for example $\beta^i = 0$), equations~(\ref{gdot2})
and~(\ref{Adot2}) therefore reduce to a coupled set of nonlinear,
inhomogeneous wave equations for the conformal metric $\tilde
\gamma_{ij}$, in which the gauge terms $K$ and $\tilde \Gamma^i$, the
conformal factor $\exp(\phi)$, and the matter terms $M_{ij}$ appear as
sources.  Wave equations not only reflect the hyperbolic nature of
general relativity, but can also be implemented numerically in a
straight-forward and stable manner.  The same method has often been
used to reduce the four-dimensional Ricci tensor
$R_{\alpha\beta}$~\cite{c62} and to bring Einstein's equations into a
symmetric hyperbolic form~\cite{fm72}.

Note that the connection functions $\tilde \Gamma^i$ are pure gauge
quantities in the sense that they could be chosen, for example, to
vanish by a suitable choice of spatial coordinates (``conformal
three-harmonic coordinates'', compare~\cite{sy78}).  The $\tilde
\Gamma^i$ would then play the role of ``conformal gauge source
functions'' (compare~\cite{c62,fm72}). Here, however, we impose the
gauge by choosing the shift $\beta^i$, and evolve the $\tilde
\Gamma^i$ with equation~(\ref{Gammadot2}) below.  
Similarly, $K$ is a pure gauge variable (and could be chosen to vanish
by imposing maximal time slicing).

An evolution equation for the $\tilde \Gamma^i$ can be
derived by permuting a time derivative with the space derivative
in~(\ref{cgsf})
\begin{eqnarray} 
\frac{\partial}{\partial t} \tilde \Gamma^i
& = & - \frac{\partial}{\partial x^j} \Big( 2 \alpha \tilde A^{ij} 
	\nonumber \\[1mm]
& &	- 2 \tilde \gamma^{m(j} \beta^{i)}_{~,m}
	+ \frac{2}{3} \tilde \gamma^{ij} \beta^l_{~,l} 
	+ \beta^l \tilde \gamma^{ij}_{~~,l} \Big).
\end{eqnarray}
It turns out to be essential for the numerical stability of the system
to eliminate the divergence of $\tilde A^{ij}$ with the help of the
momentum constraint~(\ref{mom1}), which yields
\begin{eqnarray} \label{Gammadot2}
\frac{\partial}{\partial t} \tilde \Gamma^i
& = & - 2 \tilde A^{ij} \alpha_{,j} + 2 \alpha \Big(
	\tilde \Gamma^i_{jk} \tilde A^{kj} -  \nonumber \\[1mm]
& &	\frac{2}{3} \tilde \gamma^{ij} K_{,j}
	- \tilde \gamma^{ij} S_j + 6 \tilde A^{ij} \phi_{,j} \Big) +
	\nonumber \\[1mm]
& &	\frac{\partial}{\partial x^j} \Big(
	\beta^l \tilde \gamma^{ij}_{~~,l} 
	- 2 \tilde \gamma^{m(j} \beta^{i)}_{~,m}
	+ \frac{2}{3} \tilde \gamma^{ij} \beta^l_{~,l} \Big).
\end{eqnarray}

We now consider $\phi$, $K$, $\tilde \gamma_{ij}$, $\tilde A_{ij}$ and
$\tilde \Gamma^i$ as fundamental variables.  These can be evolved with
the evolution equations~(\ref{phidot2}), (\ref{Kdot2}), (\ref{gdot2}),
(\ref{Adot2}), and~(\ref{Gammadot2}), which we call System II.  Note
that obviouly not all these variables are independent.  In particular,
the determinant of $\tilde \gamma_{ij}$ has to be unity, and the trace
of $\tilde A_{ij}$ has to vanish.  These conditions can either be used
to reduce the number of evolved quantities, or, alternatively, all 
quantities can be evolved and the conditions can be used as a numerical
check (which is what we do in our implementation).


\section{NUMERICAL IMPLEMENTATION}
\label{sec3}

In order to compare the properties of Systems I and II, we implemented
them numerically in an identical environment.  We integrate the
evolution equations with a two-level, iterative Crank-Nicholson
method.  The iteration is truncated after a certain accuracy has been
achieved.  However, we iterate at least twice, so that the scheme is
second order accurate.  

The gridpoints on the outer boundaries are updated with a Sommerfeld
condition.  We assume that, on the outer boundaries, the fundamental
variables behave like outgoing, radial waves
\begin{equation}
Q(t,r) = \frac{G(\alpha t - e^{2 \phi} r)}{r}.
\end{equation}
Here $Q$ is any of the fundamental variables (except for the diagonal
components of $\tilde \gamma_{ij}$, for which the radiative part is $Q
= \tilde \gamma_{ii} - 1$), and $G$ can be found by following the
characteristic back to the previous timestep and interpolating the
corresponding variable to that point (see also~\cite{sn95}).  We found
that a linear interpolation is adequate for our purposes.

We impose octant symmetry in order to minimize the number of
gridpoints, and impose corresponding symmetry boundary conditions on
the symmetry plains.  Unless noted otherwise, the calculations
presented in this paper were performed on grids of $(32)^3$
gridpoints, and used a Courant factor of $1/4$.  The code has been
implemented in a parallel environment on SGI Power ChallengeArray and
SGI CRAY Origin2000 computer systems at NCSA using DAGH~\cite{dagh}
software for parallel processing.


\section{RESULTS}
\label{sec4}

\subsection{Initial Data}

For initial data, we choose a linearized wave solution (which is 
then evolved with the full nonlinear systems I and II).  
Following Teukolsky~\cite{t82}, we construct a time-symmetric,
even-parity $L=2$, $M = 0$ solution.  The coefficients $A, B$ and $C$
(see equation (6) in~\cite{t82}) are derived from a function
\begin{equation}
F(t,r) = {\cal A}\, (t \pm r) \, \exp(- \lambda (t \pm r)^2).
\end{equation}
Unless noted otherwise, we present results for an amplitude
${\cal A} = 10^{-3}$ and a wavelength $\lambda = 1$.  The outer
boundary conditions are imposed at $x, y, z = 4$. 

We evolve these initial data for zero shift
\begin{equation}
\beta^i = 0,
\end{equation}
and compare the performance of Systems I and II for both geodesic and
harmonic slicing.

\subsection{Geodesic Slicing}

\begin{figure}
\epsfxsize=3in
\begin{center}
\leavevmode
\epsffile{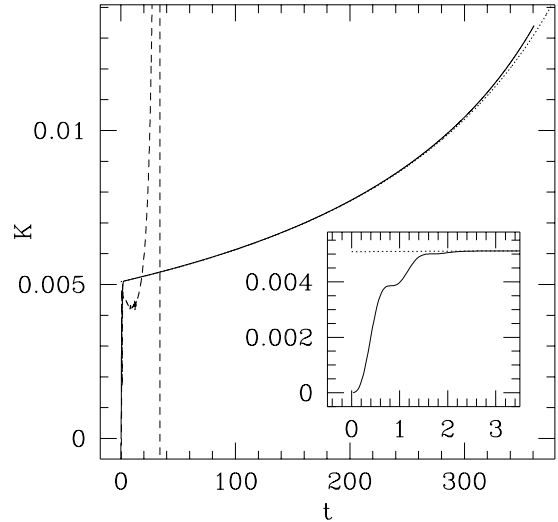}
\end{center}
\caption{Evolution of the trace of the extrinsic curvature $K$ 
for a small amplitude wave in geodesic slicing at the origin (see text
for details).  The solid line is the result for System II, and the
dashed line for System I.  The dotted line is the approximate 
solution~(\protect\ref{K_approx}).}
\end{figure}

In geodesic slicing, the lapse is unity
\begin{equation}
\alpha = 1.
\end{equation}
Since the acceleration of normal observers satisfies
$a_a = D_a \ln \alpha = 0$,
these observers follow geodesics.  The energy content of even
a small, linear wave packet will therefore focus these observers, and
even after the wave has dispersed, the observers will continue to
coast towards each other.  Since $\beta^i = 0$, normal observers are
identical to coordinate observers, hence geodesic slicing will
ultimately lead to the formation of a coordinate singularity even for
arbitrarily small waves.

\begin{figure}
\epsfxsize=3in
\begin{center}
\leavevmode
\epsffile{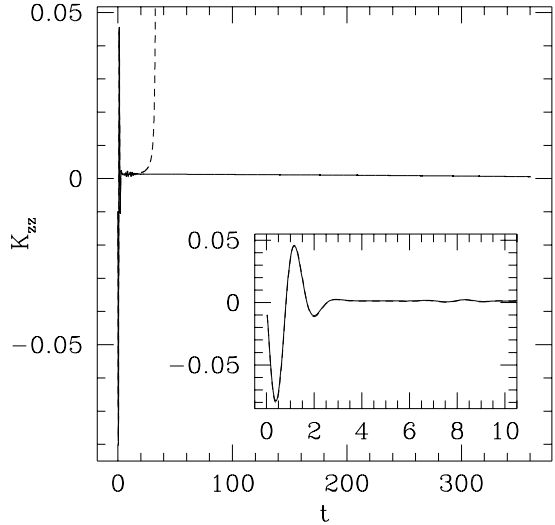}
\end{center}
\caption{Evolution of the extrinsic curvature component $K_{zz}$ at
the origin in geodesic slicing.  The solid line is the result for
System II, and the dashed line for System I.  For System II, we
constructed $K_{zz}$ from $\tilde A_{zz}$, $\phi$, $K$ and $\tilde
\gamma_{zz}$.}
\end{figure}

The timescale for the formation of this singularity can be estimated
from equation~(\ref{Kdot2}) with $\alpha = 1$ and $\beta^i = 0$.  The
$\tilde A_{ij}$, which can be associated with the gravitational waves,
will cause $K$ to increase to some finite value, say $K_0$ at time
$t_0$, even if $K$ was zero initially.  After
roughly a light crossing time, the waves will have dispersed, and the
further evolution of $K$ is described by $\partial_t K \sim K^2/3$, or
\begin{equation} \label{K_approx}
K \sim \frac{3 K_0}{3 - K_0(t - t_0)}
\end{equation}
(see~\cite{sn95}).  Obviously, the coordinate singularity forms at
$t \sim 3/K_0 + t_0$ as a result of the nonlinear evolution.  

We can now evolve the wave initial data with Systems I and II and
compare how well they reproduce the formation of the coordinate
singularity.

In Figure 1, we show $K$ at the origin ($x = y = z = 0$) as a function
of time both for System I (dashed line) and System II (solid line).
We also plot the approximate analytic solution~(\ref{K_approx}) as a
dotted line, which we have matched to the System I solution with
values $K_0 = 0.00518$ and $t_0 = 10$.  For these values,
equation~(\ref{K_approx}) predicts that the coordinate singularity
appears at $t \sim 590$.  In the insert, we show a blow-up of System
II for early times.  It can be seen very clearly how the initial wave
content lets $K$ grow from zero to the ``seed'' value $K_0$.  Once the
waves have dispersed, System II approximately follows the
solution~(\ref{K_approx}) up to fairly late times.  System I, on the
other hand, crashes long before the coordinate singularity appears.

In Figure 2, we compare the extrinsic curvature component $K_{zz}$
evaluated at the origin.  The noise around $t \sim 8$, which is
present in the evolutions of both systems, is caused by reflections of
the initial wave off the outer boundaries.  It is obvious from these
plots that System II evolves the equations stably to a fairly late
time, at which the integration eventually becomes inaccurate as the
coordinate singularity approaches.  We stopped this calculation when
the iterative Crank-Nicholson scheme no longer converged after a
certain maximum number of iterations.  It is also obvious that System
I performs extremely poorly, and crashes at a very early time, well
before the coordinate singularity.

It is important to realize that the poor performance of System I is
{\em not} an artifact of our numerical implementation.  For example,
the ADM code currently being used by the Black Hole Grand Challenge
Alliance, is based on the equations of System I, and also crashes
after a very similar time~\cite{r98} (see also~\cite{aetal98}, where a
run with a much smaller initial amplitude nevertheless crashes 
earlier than our System II).  This shows that the code's crashing is
intrinsic to the equations and slicing, and not to our numerical
implementation.

\subsection{Harmonic Slicing}

\begin{figure}
\epsfxsize=3in
\begin{center}
\leavevmode
\epsffile{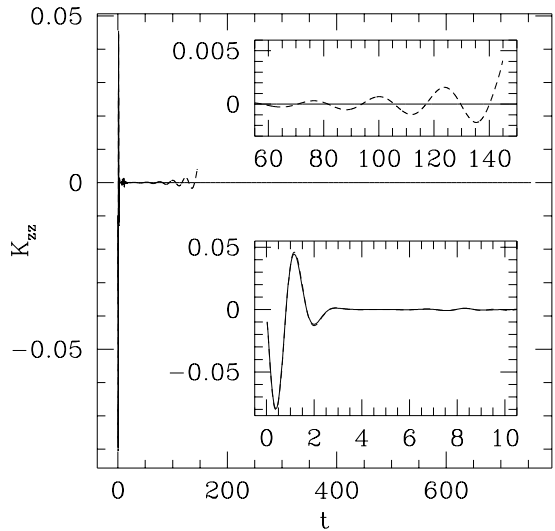}
\end{center}
\caption{Evolution of the extrinsic curvature component $K_{zz}$ at
the origin in harmonic slicing.  The solid line is the result for
System II, and the dashed line for System I.  For System II, we
constructed $K_{zz}$ from $\tilde A_{zz}$, $\phi$, $K$ and $\tilde
\gamma_{zz}$.}
\end{figure}

Since geodesic slicing is known to develop coordinate singularities
for generic, nontrivial initial data, it is obviously not a very
good slicing condition.  We therefore also compare the two Systems
using harmonic slicing.  In harmonic slicing, the coordinate time $t$
is a harmonic function of the coordinates
$\nabla^{\alpha} \nabla_{\alpha} t = 0$,
which is equivalent to the condition
\begin{equation}
\Gamma^0 \equiv g^{\alpha\beta} \Gamma^0_{\alpha\beta} = 0
\end{equation}
(where the $\Gamma^{\alpha}_{\beta\gamma}$ are the connection coefficients
associated with the four-dimensional metric $g_{\alpha\beta}$).  For
$\beta^i = 0$, the above condition reduces to
\begin{equation}
\partial_t \alpha = - \alpha^2 K.
\end{equation}
Inserting~(\ref{phidot2}), this can be written as 
\begin{equation}
\partial_t (\alpha e^{-6 \phi}) = 0
\mbox{~~~or~~~}
\alpha = C(x^i) e^{6 \phi},
\end{equation}
where $C(x^i)$ is a constant of integration, which depends on the spatial 
coordinates only.  In practice, we choose $C(x^i) = 1$. 

In Figure 3, we show results for the same initial data as in the last
section.  Obviously, both Systems do much better for this slicing
condition.  System I crashes much later than in geodesic slicing
(after about 40 light crossing times, as opposed to about 10 for
geodesic slicing), but it still crashes.  System II, on the other
hand, did not crash after even over 100 light crossing times.  We
never encountered a growing instability that caused the code to crash.


\section{SUMMARY AND CONCLUSION}
\label{sec5}

We numerically implement two different formulations of Einstein's
field equations and compare their performance for the evolution of
linear wave initial data.  System I is the standard set of ADM
equations for the evolution of $\gamma_{ij}$ and $K_{ij}$.  In System
II, we conformally decompose the equations and introduce connection
functions.  The conformal decomposition naturally splits ``radiative''
variables from ``nonradiative'' ones, and the connection functions
are used to bring the Ricci tensor into an elliptic form.  These
changes are appealing mathematically, but also have a striking
numerical consequence: System II performs far better than System I.

It is interesting to note that most earlier axisymmetric codes
(e.g.~\cite{e84}) also relied on a decomposition similar to that of
System II.  Much care was taken to identify radiative variables and to
integrate those variables as opposed to the raw metric components.  It
is surprising that this experience was abandoned in the development of
most 3+1 codes, which integrate equations equivalent to System I.
These codes have been partly successful~\cite{cetal98}, but obvious
problems remain, as for example the inability to integrate low
amplitude waves for arbitrarily long times.  While efforts have been
undertaken to stabilize such codes with the help of appropriate outer
boundary conditions~\cite{aetal98,rar98}, our findings point to the
equations themselves as the fundamental cause of the problem, and not
to the outer boundaries.  Obviously, boundary conditions as employed
in the perturbative approach in~\cite{aetal98,rar98} or in the
characteristic approach in~\cite{bglswi96} are still needed for
accuracy -- but our results clearly suggest that they are not needed
for stability~\cite{footnote4}.

Some of the recently proposed hyperbolic systems are very appealing in
that they bring the equations in a first order, symmetric hyperbolic
form, and that all characteristics are physical (i.e., are either at
rest with respect to normal observers or travel with the speed of
light)~\cite{aacy96,acy98}.  These properties may be very advantageous
for numerical implementations, in particular at the boundaries (both
outer boundaries and, in the case of black hole evolutions, inner
``apparent horizon'' boundaries). Some of these systems have also been
implemented numerically, and show stability properties very similar to
our System II~\cite{cs98}.  Our System II, on the other hand, uses
fewer variables than most of the hyperbolic formulations, and
does not take derivatives of the equations, which may be advantageous
especially when matter sources are present.  This suggests that a
system similar to System II may be a good choice for evolving interior
solutions and matter sources, while one may want to match to one of
the hyperbolic formulations for a better treatment of the boundaries.

The mathematical structure of System II is more appealing than that of
System I, and these improvements are reflected in the numerical
behavior.  We therefore conclude that the mathematical structure has a
very deep impact on the numerical behavior, and that the ability to
finite difference the standard ``$\dot g - \dot K$'' ADM equations may
not be sufficient to warrant a stable evolution.

\acknowledgments

It is a pleasure to thank A.~M.~Abrahams, L.~Rezzolla, J.~W. York and
M.~Shibata for many very useful conversations.  We would also like to
thank H.~Friedrich for very valuable comments, and S.~A.~Hughes for a
careful checking of our code.  Calculations were performed on SGI CRAY
Origin2000 computer systems at the National Center for Supercomputing
Applications, University of Illinois at Urbana-Champaign.  This work
was supported by NSF Grant AST 96-18524 and NASA Grant NAG 5-3420 at
Illinois, and the NSF Binary Black Hole Grand Challenge Grant Nos. NSF
PHYS 93-18152, NSF PHY 93-10083 and ASC 93-18152 (ARPA supplemented).



\begin{references}
\bibitem{footnote1} Note, however, that Arnowitt, Deser and 
	Misner~\cite{adm62} wrote these equations in terms of the
	conjugate momenta $\pi_{ij}$ instead of the extrinsic 
	curvature $K_{ij}$.
\end{references}
\end{document}